\documentstyle[12pt]{article}

\catcode`\@=11
\long\def\@makefntext#1{
\protect\noindent \hbox to 3.2pt {\hskip-.9pt
$^{{\ninerm\@thefnmark}}$\hfil}#1\hfill}		

\def\@makefnmark{\hbox to 0pt{$^{\@thefnmark}$\hss}}  

\def\ps@myheadings{\let\@mkboth\@gobbletwo
\def\@oddhead{\hbox{}
\rightmark\hfil\ninerm\thepage}
\def\@oddfoot{}\def\@evenhead{\ninerm\thepage\hfil
\leftmark\hbox{}}\def\@evenfoot{}
\def\sectionmark##1{}\def\subsectionmark##1{}}

\renewcommand{\thefootnote}{\fnsymbol{footnote}}

\newcounter{sectionc}\newcounter{subsectionc}\newcounter{subsubsectionc}
\renewcommand{\section}[1] {\vspace*{0.6cm}\addtocounter{sectionc}{1}
\setcounter{subsectionc}{0}\setcounter{subsubsectionc}{0}\noindent
	{\normalsize\bf\thesectionc. #1}\par\vspace*{0.4cm}}
\renewcommand{\subsection}[1] {\vspace*{0.6cm}\addtocounter{subsectionc}{1}
	\setcounter{subsubsectionc}{0}\noindent
	{\normalsize\it\thesectionc.\thesubsectionc. #1}\par\vspace*{0.4cm}}
\renewcommand{\subsubsection}[1]
{\vspace*{0.6cm}\addtocounter{subsubsectionc}{1}
\noindent {\normalsize\rm\thesectionc.\thesubsectionc.\thesubsubsectionc.
	#1}\par\vspace*{0.4cm}}

\newcounter{appendixc}
\newcounter{subappendixc}[appendixc]
\newcounter{subsubappendixc}[subappendixc]

\renewcommand{\appendix}[1] {\vspace*{0.6cm}
        \refstepcounter{appendixc}
        \setcounter{figure}{0}
        \setcounter{table}{0}
        \setcounter{equation}{0}
        \renewcommand{\thefigure}{\Alph{appendixc}.\arabic{figure}}
        \renewcommand{\thetable}{\Alph{appendixc}.\arabic{table}}
        \renewcommand{\theappendixc}{\Alph{appendixc}}
        \renewcommand{\theequation}{\Alph{appendixc}.\arabic{equation}}
        \noindent{\bf Appendix \theappendixc #1}\par\vspace*{0.4cm}}

\def\abstracts#1{{
\centering{\begin{minipage}{12.2truecm}\footnotesize\baselineskip=12pt\noindent
	\centerline{\footnotesize ABSTRACT}\vspace*{0.3cm}
	\parindent=0pt #1
	\end{minipage}}\par}}

\renewenvironment{thebibliography}[1]
	{\begin{list}{\arabic{enumi}.}
	{\usecounter{enumi}\setlength{\parsep}{0pt}
\setlength{\leftmargin 1.25cm}{\rightmargin 0pt}
	 \setlength{\itemsep}{0pt} \settowidth
	{\labelwidth}{#1.}\sloppy}}{\end{list}}

\newcounter{itemlistc}
\newcounter{romanlistc}
\newcounter{alphlistc}
\newcounter{arabiclistc}

\newcommand{\fcaption}[1]{
        \refstepcounter{figure}
        \setbox\@tempboxa = \hbox{\footnotesize Fig.~\thefigure. #1}
        \ifdim \wd\@tempboxa > 6in
           {\begin{center}
        \parbox{6in}{\footnotesize\baselineskip=12pt Fig.~\thefigure. #1}
            \end{center}}
        \else
             {\begin{center}
             {\footnotesize Fig.~\thefigure. #1}
              \end{center}}
        \fi}

\newcommand{\tcaption}[1]{
        \refstepcounter{table}
        \setbox\@tempboxa = \hbox{\footnotesize Table~\thetable. #1}
        \ifdim \wd\@tempboxa > 6in
           {\begin{center}
        \parbox{6in}{\footnotesize\baselineskip=12pt Table~\thetable. #1}
            \end{center}}
        \else
             {\begin{center}
             {\footnotesize Table~\thetable. #1}
              \end{center}}
        \fi}

\def\@citex[#1]#2{\if@filesw\immediate\write\@auxout
	{\string\citation{#2}}\fi
\def\@citea{}\@cite{\@for\@citeb:=#2\do
	{\@citea\def\@citea{,}\@ifundefined
	{b@\@citeb}{{\bf ?}\@warning
	{Citation `\@citeb' on page \thepage \space undefined}}
	{\csname b@\@citeb\endcsname}}}{#1}}

\newif\if@cghi
\def\cite{\@cghitrue\@ifnextchar [{\@tempswatrue
	\@citex}{\@tempswafalse\@citex[]}}
\def\citelow{\@cghifalse\@ifnextchar [{\@tempswatrue
	\@citex}{\@tempswafalse\@citex[]}}
\def\@cite#1#2{{$\null^{#1}$\if@tempswa\typeout
	{IJCGA warning: optional citation argument
	ignored: `#2'} \fi}}

 1
 1
 1

\font\ninerm=cmr9

\begin{document}

\newcommand{\st}{\scriptstyle}
\newcommand{\sst}{\scriptscriptstyle}
\newcommand{\mco}{\multicolumn}
\newcommand{\epp}{\epsilon^{\prime}}
\newcommand{\vep}{\varepsilon}
\newcommand{\ra}{\rightarrow}
\newcommand{\ppg}{\pi^+\pi^-\gamma}
\newcommand{\vp}{{\bf p}}
\newcommand{\ko}{K^0}
\newcommand{\kb}{\bar{K^0}}
\newcommand{\al}{\alpha}
\newcommand{\ab}{\bar{\alpha}}
\def\be{\begin{equation}}
\def\ee{\end{equation}}
\def\bea{\begin{eqnarray}}
\def\eea{\end{eqnarray}}
\def\CPbar{\hbox{{\rm CP}\hskip-1.80em{/}}}

March, 1995 \hfill LBL-36910\\
\vskip 0.2in
\centerline{\normalsize\bf DO $W_L$ AND $H$ FORM A P-WAVE BOUND
STATE?\footnote{This
work was supported
by the Director, Office of Energy
Research, Office of High Energy and Nuclear Physics,
Divisions of High
Energy Physics
 of the U.S. Department of Energy under Contract
DE-AC03-76SF00098
 and by the Natural Sciences and
Engineering Research Council of Canada.}
}
\baselineskip=16pt
\centerline{\footnotesize ZHENG HUANG\footnote{Talk presented
at the Beyond The Standard Model IV, 13-18 December 1994, Lake Tahoe,
California}}
\baselineskip=13pt
\centerline{\footnotesize\it Theoretical Physics Group, Lawrence
Berkeley Laboratory}
\baselineskip=12pt
\centerline{\footnotesize\it University of California, Berkeley, CA 94720, USA}
\centerline{\footnotesize E-mail: huang@theorm.lbl.gov}

\vspace*{0.5cm}
\abstracts{We examine the possibility of bound state formation in the
$W_LH\rightarrow W_LH$ channel. The dynamical calculation using the
$N/D$ method indicates that when the interactions among the Goldstone and
Higgs bosons become sufficiently strong, a $p$-wave state
$[I^G(J^P)=1^-(1^+)]$ may emerge.}

\vspace*{0.5cm}
\setcounter{footnote}{0}
\renewcommand{\thefootnote}{\alph{footnote}}
We shall consider the elastic scattering of  $W_L H$
and view the process as the
dynamical force for the possible generation of bound states or
resonances, of which $W_L$ and $H$ are constituents.
To study the $W_L H$ scattering at high energies, it is much simpler to
work with the Goldstone bosons ($w^\pm,z$) and $H$
by invoking the Equivalence Theorem \cite{eq,lqt,chanogail}
when away from the threshold.
We assume that the electroweak symmetry-breaking sector can be
effectively parameterized by the linear $\sigma$-model.
%
%
It is sufficient to consider the $I_3=0$ channel $zH\rightarrow
zH$ ($w^\pm H$ are similar).
When gauge interactions are ignored, it
is isolated and decoupled from other strong scattering channels.
The Born amplitude for $zH\rightarrow zH$ is
\begin{equation}
{\cal T}^B(s,t,u)=-2i\lambda \left[ 1+\frac{m^2_H}{s-M_Z^2}+
\frac{3m_H^2}{t-m_H^2}+\frac{m_H^2}{u-M_Z^2}\right]\; , \label{tb}
\end{equation}
where  $m_H^2=2\lambda\upsilon^2$ and $\upsilon =246$ GeV;
$s$, $t$ and $u$ are the Mandelstam variables:
$t = -2\nu(1-\cos\theta )$ and
$u =(M_Z^2+s_0-s)+2\nu(1-\cos\theta)$,
where $s_0=2m_H^2+M_Z^2$ and $\nu$ is the CM momentum square.

Before we go on and discuss the dynamical feature of this scattering amplitude,
some special attention  has to be paid to the $u$-channel $z$-exchange.
The matrix element is formally divergent at some scattering angle
$ \cos\theta =1+(s_0-s)/(2\nu)$
when $(m_H+M_Z)^2<s_0\leq s$, at which $u-M_Z^2=0$.
This is only possible when $m_H>2M_Z$ and $s\geq s_0$.
The first inequality is satisfied when $m_H>2M_Z$ which  implies
that $H$ is necessarily unstable. The nature of the singularity
is logarithmic and can be seen in the $p$-wave amplitude
\begin{eqnarray}
a_1^B  &=&
 \frac{-\lambda}{16\pi}\left[ \frac{2m_H^2}{2\nu}+
\frac{3m_H^2(2\nu+m_H^2)}{4\nu^2}\ln\frac{m_H^2}{4\nu +m_H^2}
\right. \nonumber\\
 & &\left.
-\frac{m_H^2}{4\nu^2}(2\nu +s_0-s)
\ln \frac{s_0-s}{4\nu +s_0-s} \right] \; .
\end{eqnarray}
The
amplitude goes to negative infinity at the location of the singularity
$s = s_0$, which seemingly represents a repulsive force.
However, such a singularity is superficial.
The origin of such a singularity can be traced back to the inconsistent
treatment of the unstable Higgs boson: $H$ could first decay
into two $z$'s, one of which can subsequently combine the other
initial state $z$ into $H$ again when $s\geq s_0$.
The $u$-channel $z$-exchange thus represents a real process
(actually two successive  real processes)
and the total cross section is formally divergent.
The solution to this superficial singularity
lies precisely on the fact that the Higgs boson is unstable.
The logarithmic singularity is smeared by the large uncertainty in the
Higgs mass position due to the finite width of the Higgs boson,
and effectively the singularity does not exist.
Our minimal prescription is to allow the Higgs boson to develop a complex
energy due to the Higgs width ($\Gamma_H$),  but nevertheless to
retain the quasi-two-body structure of the amplitude (a more strict
treatment would be the consideration
The ``on-shell'' condition for  an  unstable Higgs boson is then
\begin{equation}
(E_H-i\omega )^2-\nu=m_H^2-im_H\Gamma_H\; ,\label{width}
\end{equation}
where $E_H$ ($\omega$) is the real (imaginary)
part of the energy in the CM system,
and $\Gamma_H$ the Higgs decay width.
The modified Mandelstam variables $\hat{s}$, $\hat{t}$
and $\hat{u}$, first introduced by Peierls \cite{peierls}, are
\begin{eqnarray}
\hat{s} =  s-im_H\Gamma_H\left( 1+\frac{E_Z}{E_H}\right);\quad  \quad
\hat{t}  =  -2\nu(1-\cos\theta );\\
\hat{u}  =  (M_Z^2 + s_0-s)+2\nu(1-\cos\theta )
-im_H\Gamma_H\left( 1-\frac{E_Z}{E_H}\right)
\end{eqnarray}
where $s$ is redefined as
$(E_Z+E_H)^2- {m_H^2\Gamma^2_H}/{4E_H^2}$.
Note that  the threshold value of $s$ is
$s_{\rm th} \equiv s(\nu=0)$
which is smaller than $(m_H+M_Z)^2$ for the unstable $H$.
With these modifications, the partial wave  amplitude becomes
completely regular
and the principal part of the $p$-wave Born
amplitude is positive  over the whole physical region, thus representing an
attractive  force.

We now examine
whether the interaction may provide a dynamical driving force
strong enough to lead  to the formation of
bound states. Our goal is to sum up a class of ladder
diagrams, according to the requirement of  unitarity.
This is done most consistently by an $N/D$ method in dispersion theory
\cite{nd}.  The full partial wave amplitude $a_1(s)$
must satisfy the elastic unitarity condition:
${\rm Im}a_1(s)=-\sqrt{ 4\nu/s }\; a_1^*a_1  \quad (s>s_{\rm th})$.
In the $N/D$ method, $a_1(s)$ is written as $N(s)/D(s)$ where $N(s)$ has
only left-hand cuts and $D(s)$ has only right-hand cuts. An once-subtracted
dispersion relation may be written for $D(s)$
\begin{equation}
D(s)=1-\frac{(s-\mu^2)}{\pi}\int_{s_{\rm th}}^{\infty}ds'
\sqrt{\frac{4\nu (s')}{s}}\frac{N(s')}{(s'-\mu^2)(s'-s)}
\end{equation}
where $\mu$ is the subtraction point chosen to be at
the reduced mass of the system.
Unlike the ``bootstrap'' approach where the bound state itself should
be included in the cross channels,
$N(s)$ is approximated by the principal part of the Born amplitude
(\ref{tb})
involving only elementary fields ($z$ and $H$) for the calculation
of a loose bound state.
If for some value $s=s_B$ ($0<s_B<s_{\rm th}$), $D(s)$ vanishes, it implies
that the scattering amplitude $a_1(s)$ has a pole at $s_B$ which can be
interpreted as the mass location of a bound state.

Our numerical calculation shows that $D(s)$ for the $p$-wave
develops a zero only when
$m_H \geq 1$ TeV which coincides with a
well-known unitarity bound first obtained by
Lee, Quigg and Thacker \cite{lqt}.
We shall call this bound state ${A}_1$ as opposed to the QCD counterpart
which  is now called $a_1$.
As a result of strong self-coupling, some
interesting particle spectrum besides the Higgs boson may emerge. In our model,
the $u$-channel $z$-exchange turns out to be very important since we also
checked the result without the $u$-channel contribution and found no bound
states.
As for the $s$-wave, the presence of the $s$-channel contribution
provides extra repulsive force so that  bound states do not form.
In Fig.\ \ref{masses}, we show the calculated  mass of $A_1$, $M_A^{}$,
versus the parameter $m_H^{}$.
The binding energy
$B_A^{}  \equiv (m_H+M_Z)- M_A$ is only modest
about order of $M_Z$ for $m_H\sim 1.5$ TeV.
The linearity of the curve represents the fact  that the binding energy
is proportional to  the square root of the strength of the
self-coupling $\lambda$. The primary decay modes for such a bound state
would be  $W_L + (2W_L)_{\rm s-wave}$ through
a Higgs boson exchange. The width can, in principle, be determined from the
coupling $g_{AZH}^{}$ which can be calculated from the residue of
$a_1(s)$ at $s=M_A^2$. Such an axial vector state contributes to a negative
value of the $S$ parameter in the precision measurement. More
detailed implications can be found in Ref.\ \cite{hhh}.

\section{Acknowledgements}
I wish to thank T.\ Han and P.Q.\ Hung for collaboration on this subject.

\begin{figure}
\center
\vskip 1in
\fcaption{Calculated mass $M_A^{}$ versus the input parameter $m_H$.
For comparison, $m_H^{}+M_Z^{}$ is also presented by the dotted line.}
\label{masses}
\end{figure}

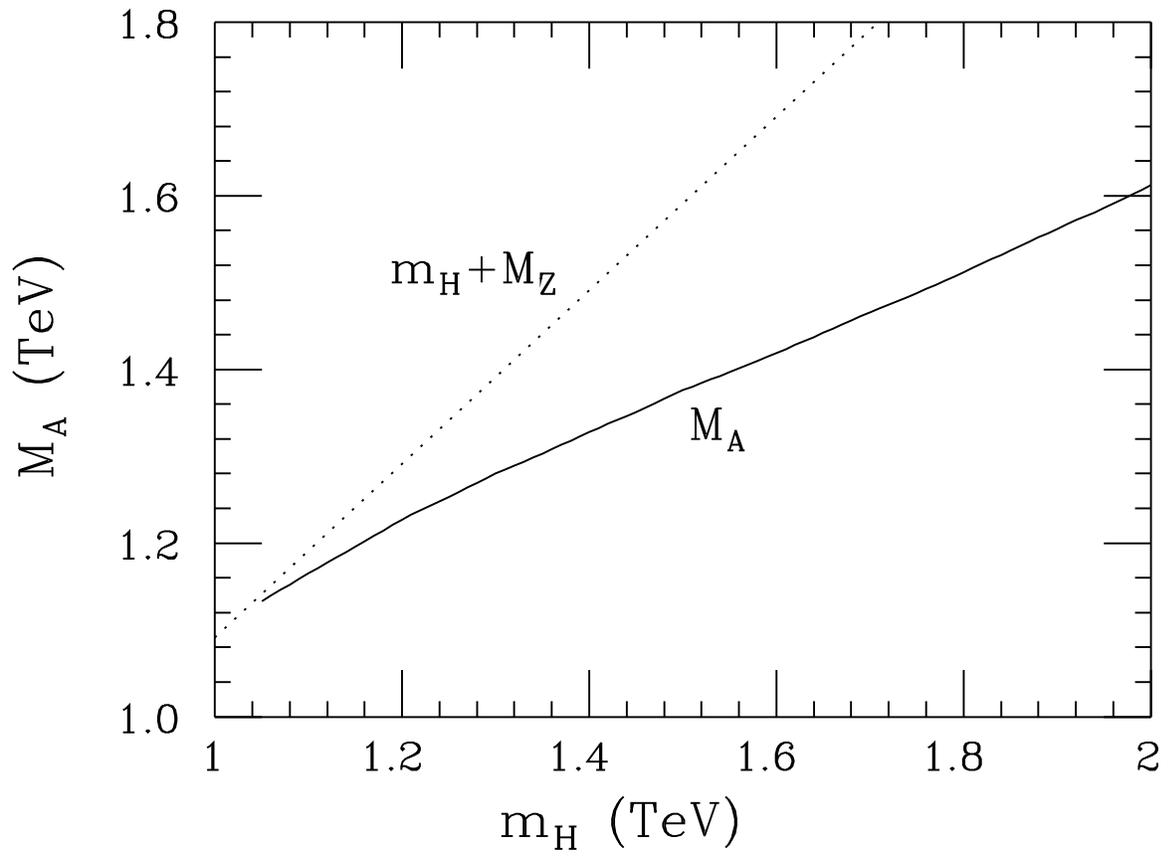
\end{document}